\magnification 1200
\centerline {\bf On Order, Disorder and Coherence}
\vskip 0.5cm
\centerline {\bf by Geoffrey Sewell}
\vskip 0.5cm
\centerline {\bf Department of Physics, Queen Mary, University of London,}
\vskip 0.2cm
\centerline {\bf Mile End Road, London E1 4NS, UK}
\vskip 1cm
\centerline {\bf Abstract}
\vskip 0.3cm
We provide a brief survey of  quantum statistical characterisations of order, disorder and coherence in 
systems of many degree of freedom. Here, order and coherence are described in terms of symmetry 
breakdown, while disorder is described in terms of entropy and algorithmic complexity, whose 
interconnection has been recently extended from the classical to the quantum domain. We see that, in the 
present physical context, the concepts of order and disorder are not mutually antithetical but bear an 
interrelationship similar to that between signals and noise. 
\vskip 0.5cm
\centerline {\bf 1. Introduction.}
\vskip 0.3cm
This article is designed to provide a succinct account of the prevailing quantum statistical pictures of order 
and disorder in systems with many degrees of freedom. In this context, order essentially signifies 
organisation of the microscopic components of such systems to produce macroscopic fields, or signals, as 
exemplified by the polarisation of a ferromagnet; while disorder amounts to randomness.
\vskip 0.2cm
Here we formulate mathematical pictures of order and disorder within the framework of operator algebraic 
statistical mechanics [Em, Th, Se1], which provides a natural setting for their descriptions. We start in Sec. 
2 with a brief sketch of the structure of algebraic quantum theory. We then pass on, in Sec. 3, to both the 
probabilistic formulation of  disorder in terms of Von Neumann\rq s entropy [VN] and the intrinsic 
description thereof by Kolmogorov\rq s algorithmic complexity [Ko].  In particular, we discuss Brudno\rq s 
theorem [Br] and its recent quantum generalisation [BKMSS], which shows that these two characterisations 
of disorder essentially yield the same picture. In Sec. 4, we formulate the concept of order due to symmetry 
breakdown. In Sec. 5, we refine this formulation of order to an extreme version thereof, namely coherence, 
in the sense proposed by Glauber [Gl]. We provide some concrete examples both of order, in Sec. 4, and of 
coherence in Sec. 5. We conclude in Sec. 6 with some further brief observations about order, disorder and 
coherence, and discuss the need to widen the concept of order to the description of organisational structures 
that are not covered by existing theories. 
\vskip 0.5cm
\centerline {\bf 2. The Operator Algebraic Framework}
\vskip 0.3cm
We employ the standard operator algebraic description [Em, Th, Se1] of a quantum  mechanical system, 
${\Sigma}$, as a triple $({\cal A}, {\cal S},{\alpha})$ representing its observables, states and dynamics, 
respectively. Specifically, ${\cal A}$ is a $C^{\star}$-algebra, whose self-adjoint elements represent the 
bounded observables of ${\Sigma}$, and ${\alpha}$ is a homomorphism of {\it either} the additive group 
${\bf R}$ into the automorphisms of ${\cal A}$ {\it or} of the semigroup ${\bf R}_{+}$ into completely 
positive, identity preserving, linear contractions of this algebra, according to whether ${\Sigma}$ is a 
conservative system or an open Markovian dissipative one. The state space ${\cal S}$ is a norm closed, 
convex subset of the positive, normalised, linear functionals on ${\cal A}$ that is stable under the action of 
the dual of ${\alpha}$. We shall denote the expectation value of $A \ ({\in}{\cal A})$ for the state 
${\rho}$ by ${\rho}(A)$ or, equivalently, ${\langle}{\rho};A{\rangle}$.The pure states are the extremal 
elements of ${\cal S}$.Thus, the model is specified by the structures of ${\cal A}, \ {\cal S}$ and 
${\alpha}$.  We note that this generic model also covers the case of classical mechanical systems, which 
are distinguished by the condition that ${\cal A}$ is abelian. In this case, by the Gelfand isomorphism, 
${\cal A}$ is the algebra of continuous functions on a compact space $K, \ {\cal S}$ is a set of probability 
measures on $K$ and the transformations ${\alpha}_{t}$ are implemented by transformations ${\tau}_{t}$ 
of $K$. Here $K$ is the \lq phase space\rq\ of the model.  
\vskip 0.3cm
{\bf The Finite System Model} [VN].  For this,  ${\cal A}$ is the $W^{\star}$-algebra of bounded 
operators in a Hilbert space ${\cal H}$ and ${\cal S}$ is the set of  normal, i.e. ultraweakly continuous, 
states, ${\rho}$, on ${\cal A}$: these correspond to density matrices, denoted by the same symbol, 
according to the formula ${\rho}(A){\equiv}Tr({\rho}A)$. Thus ${\cal S}$ is a convex set and its extremal 
elements, representing the pure states, are those whose density matrices are one-dimensional projectors. 
In the case where ${\Sigma}$ is conservative, its dynamical automorphisms, ${\alpha}_{t}$, are unitary 
transformations $A{\rightarrow}{\rm exp}(iHt)A{\exp}(-iHt)$ of ${\cal A}$, where $H$ is the 
Hamiltonian operator of ${\Sigma}$ in units for which ${\hbar}=1$. In the case where the system is 
dissipative and its dynamical semigroup ${\alpha}$ is strongly continuous, its generator $L$ takes the 
following form [Li].
$$LA=i[H,A]_{-}+{\sum}_{r}\bigl(V_{r}^{\star}AV_{r}-{1\over 2}[V_{r}^{\star}V_{r},A]_{+}\bigr),
\eqno(2.1)$$
where $H \ (=H^{\star}),  \ V_{r}$ and ${\sum}_{r}V_{r}^{\star}V_{r}$ belong to ${\cal A}$ and 
$[.,.]_{\mp}$ denote commutator and anticommutator, respectively. 
\vskip 0.3cm
{\bf The Infinite System Model} [Em, Th, Se1]. This represents a system, ${\Sigma}$, of particles that 
occupies an infinitely extended space $X$, which we take to be either a Euclidean continuum, ${\bf 
R}^{d}$, or a lattice, ${\bf Z}^{d}$. We denote by ${\cal L}$ the set of all bounded open regions of $X$, 
and, for each ${\Lambda}$ in ${\cal L}$, we construct a $W^{\star}$-algebra, ${\cal A}_{\Lambda}$, of 
observables that is just that of a system, ${\Sigma}_{\Lambda}$, of particles of the given species confined 
to ${\Lambda}$. These local algebras are constructed so as to satisfy the natural demands that ${\cal 
A}_{\Lambda}$ is isotonic with respect to ${\Lambda}$ and that ${\cal A}_{\Lambda}$ and ${\cal 
A}_{{\Lambda}^{\prime}}$  intercommute if ${\Lambda}$ and ${\Lambda}^{\prime}$ are disjoint. It 
follows from the isotony property that ${\cal A}_{\cal L}:={\bigcup}_{{\Lambda}{\in}{\cal L}}{\cal 
A}_{\Lambda}$, is well-defined normed $^{\star}$-algebra. We designate its norm completion, ${\cal 
A}$, to be the $C^{\star}$-algebra of the bounded observables of ${\Sigma}$. We assume that this algebra 
is equipped with a representation, ${\gamma}$, of the space translation group $X$ in its automorphisms, 
which satsifies the covariance condition  that ${\gamma}(x){\cal A}_{\Lambda}{\equiv}{\cal 
A}_{{\Lambda}+x}$.
\vskip 0.2cm
We assume that the state space, ${\cal S}$, is a convex set of positive, normalised, linear functionals on 
${\cal A}$, whose restrictions to the local algebras ${\cal A}_{\Lambda}$ are normal: the local normality 
condition serves  to exclude the possibility of finding an infinity of particles in a bounded spatial region 
[DDR]. 
\vskip 0.2cm
The dynamics of ${\Sigma}$ is formulated as a natural infinite volume limit of that of the finite system 
${\Sigma}_{\Lambda}$. In the conservative case, this latter dynamics is governed by the form of the 
Hamiltonian operator, $H_{\Lambda}$, affiliated\footnote*{A possibly unbounded operator $Q$ in the 
representation space of a $W^{\star}$-algebra ${\cal B}$ is said to be affiliated to ${\cal B}$ if it 
commutes with ${\cal B}^{\prime}$, the commutant of ${\cal B}$.} to ${\cal A}_{\Lambda}$. Thus, if 
$A$ is an element of ${\cal A}_{\cal L}$ and therefore of ${\cal A}_{\Lambda}$ for ${\Lambda}$ 
sufficiently large, 
${\rm exp}(iH_{\Lambda}t)A{\rm exp}(-iH_{\Lambda}t)$ is  its evolute at time $t$ with respect to the 
dynamics of ${\Sigma}_{\Lambda}$. In the simplest cases, such as that of lattice systems with short range 
interactions [St, Ro], this coverges in norm to a definite limit as ${\Lambda}$ increases to $X$ over a 
sequence of suitably regular regions and thus yields  a definition of the dynamical automorphisms 
${\alpha}$ by the formula   
$${\alpha}_{t}A={\rm norm}:{\rm lim}_{{\Lambda}{\uparrow}X}
{\rm exp}(iH_{\Lambda}t)A{\rm exp}(-iH_{\Lambda}t) \ {\forall} \ A{\in}
{\cal A}_{\cal L}, \ t{\in}{\bf R}.\eqno(2.2)$$ 
More generally, when the convergence condition for this formula is not fulfilled, ${\cal S}$ has to be 
formulated so as to comprise just those states that support a limit dynamics represented by a weaker form 
of Eq. (2.2) [Se1, 2]. In the case where ${\Sigma}$ is an open dissipative system, its dynamical semigroup 
is similarly defined as a limit of that of the corresponding finite system ${\Sigma}_{\Lambda}$. 
\vskip 0.2cm
Thus the model of ${\Sigma}$, is represented by the quadruple $({\cal A}, \ {\cal S}, \ {\gamma}, \ 
{\alpha})$. The subset ${\cal S}_{X}$ of ${\cal S}$ comprising the space translationally invariant states is 
manifestly convex and we denote by ${\cal E}({\cal S}_{X})$ the set of its extremal elements. These are 
termed the {\it spatially ergodic} states. 
\vskip 0.3cm
{\bf Affiliated Quantum Fields.} Identifying the algebra ${\cal A}$ with any faithful representation 
thereof, a quantum field, ${\xi}(x)$, of the model is defined to be a distribution valued operator that is 
covariant with respect to space translations and whose integral against a test function $f(x)$ with compact 
support is affiliated to the local algebras ${\cal A}_{\Lambda}$ for which ${\Lambda}{\supset}{\rm 
supp}(f)$.
\vskip 0.3cm
{\bf Explicit Constructions.} The local algebras ${\cal A}_{\Lambda}$, the space translation group 
${\gamma}$ and the local Hamiltonians $H_{\Lambda}$, on which the model of ${\Sigma}$ is based, are 
constructed as follows (cf. [St, Ro, HHW] or the general treatments [Em, Th, Se1]). 
\vskip 0.2cm
In the case where ${\Sigma}$ is a system of particles, e.g. Pauli spins, on a lattice $X={\bf Z}^{d}$, we 
assume that the algebra of observables, ${\cal A}_{0}$, of each particle is that of the operators in a finite 
dimensional Hilbert space, ${\cal H}_{0}$. We take the local algebra ${\cal A}_{\Lambda}$, for 
${\Lambda}{\in}{\cal L}$, to be the tensor product ${\otimes}_{x{\in}{\Lambda}}{\cal A}_{x}$ of 
copies ${\cal A}_{x}$ of ${\cal A}_{0}$ attached to the respective sites $x$ in ${\Lambda}$;  and, for 
${\Lambda}{\subset}{\Lambda}^{\prime}$, we identify $A \ ({\in}{\cal A}_{\Lambda})$ with 
$A{\otimes}I_{{\Lambda}^{\prime} {\backslash}{\Lambda}} \ ({\in}{\cal A}_{{\Lambda}^{\prime}})$. 
Under this identification, the algebras ${\lbrace}{\cal A}_{\Lambda}{\vert}{\Lambda}{\in}
{\cal L}{\rbrace}$ satisfy the conditions of isotony and local commutativity and thus permit the above 
definitions of ${\cal A}_{\cal L}$ and ${\cal A}$. Further, denoting by $a_{x}$ the copy in ${\cal 
A}_{x}$ of the element $a_{0}$ of ${\cal A}_{0}$, we define the space translational automorphism group 
${\gamma}$ by the formula ${\gamma}(x)a_{x^{\prime}}=a_{x+x^{\prime}}$. Thus, the local algebras 
${\lbrace}{\cal A}_{\Lambda}{\rbrace}$ transform covariantly under this group. The local Hamiltonian 
$H_{\Lambda}$ is the element of ${\cal A}_{\Lambda}$ representing the interaction energy involving 
only the particles in ${\Lambda}$.
\vskip 0.2cm
In the case where ${\Sigma}$ is a system of particles of one species in the Euclidean continuum $X={\bf 
R}^{d}$, we formulate its observables in second quantisation, as expressed in terms of a quantised scalar 
or spinor field ${\psi}$, according to whether the particles are bosons or fermions. In either case, ${\psi}$ 
is a distribution valued operator in Fock space ${\cal H}_{F}$. Its algebraic properties are governed by the 
canonical commutation or anticommutation relations according to whether ${\Sigma}$ is composed of 
bosons or fermions. We define the local Hilbert space ${\cal H}_{\Lambda}$ to be the subspace of ${\cal 
H}_{F}$ generated by application to the Fock vacuum of the polynomials in the smeared fields obtained 
by integrating ${\psi}^{\star}$ against ${\cal D}({\Lambda})$-class test functions and we define the local 
algebra ${\cal A}_{\Lambda}$  to be that of the bounded operators in ${\cal H}_{\Lambda}$. The space 
translational automorphism group ${\gamma}$ is defined by the canonical formula 
${\gamma}(x){\psi}(x^{\prime}){\equiv}{\psi}(x+x^{\prime})$ . The local Hamiltonian $H_{\Lambda}$ 
is just that of ${\Sigma}_{\Lambda}$ and we assume that it, and consequently also the automorphisms 
${\alpha}_{t}$, is invariant under the gauge automorphisms ${\psi}(x){\rightarrow}{\psi}(x)
{\rm exp}(ic)$, with $c$ real and constant. 
\vskip 0.3cm
{\bf Equilibrium States of Conservative Systems.} The equilibrium states of a conservative system 
${\Sigma}$ at inverse temperature ${\beta}$ are characterised by the Kubo-Martin-Schwinger (KMS) 
condition, namely (cf. [HHW, Em, Th, Se1]) 
$${\langle}{\rho};[{\alpha}_{t}A]B{\rangle}={\langle}{\rho};B{\alpha}_{t+i{\beta}}A
{\rangle} \ {\forall} \ A,B{\in}{\cal A}, \ t{\in}{\bf R}.\eqno(2.3)$$
This represents various conditions of dynamical and thermodynamical stability [Se1] and, in the case of an 
infinite system, it automatically ensures that ${\rho}$ is locally normal [TW]. In general, it follows from 
Eq. (2.3) that its equilibrium (KMS) states at the inverse temperature ${\beta}$ comprise  a convex set, 
which we denote by ${\cal S}_{\beta}$. In the case where ${\Sigma}$ is a finite system,  ${\cal 
S}_{\beta}$ consists of just the canonical state with density matrix 
${\rm exp}(-{\beta}H)/{\rm Tr}({\rm Idem})$. By contrast, for  an infinite system, ${\cal S}_{\beta}$ is a  
Choquet simplex, which may contain more than one element and whose decomposition into extremals is 
just the central one [Ru, EKV]. Thus the set ${\cal E}({\cal S}_{\beta})$, of its extremals consists of 
primary states and may naturally be interpreted as comprising the pure equilibrium phases [EKV] of the 
system. Moreover, as they are primary, they enjoy the clustering property that [Ru]
$${\rm lim}_{{\vert}x{\vert}\to\infty}\bigl[{\langle}{\rho};A{\gamma}(x)B{\rangle}-
{\langle}{\rho};A{\rangle}{\langle}{\rho};{\gamma}(x)B{\rangle}\bigr]=0 \ {\forall} \ A,B{\in}
{\cal A}.\eqno(2.4)$$
\vskip 0.3cm
{\bf Open Dissipative Systems.} The situation is different for these systems since they carry no natural 
counterpart of the KMS states. In particular, the model $({\cal A},{\cal S}, {\gamma},{\alpha})$ does not 
necessarily have any stationary, dynamically stable primary states, which could be the counterparts of the 
pure phase equilibrium states of conservative systems. For example, as we shall discuss in Sec. 5, the stable 
primary states of a laser model, for a certain range of values of its parameters, are period functions of time 
[HL1, AS].
\vskip 0.5cm
\centerline {\bf 3. Entropy, Algorithmic Complexity and Disorder.}
\vskip 0.3cm
{\bf Entropy and Disorder.} The entropy, $S({\rho})$, of a state ${\rho}$ is given by Von Neumann\rq s 
formula [VN], which, in units for which Boltzmann\rq s constant is ${\rm log}_{2}(e)$, takes the 
following form. 
$$S({\rho})=-{\rm Tr}\bigl({\rho}{\rm log}_{2}({\rho})\bigr).\eqno(3.1)$$
In order to expose the probabilistic character of this formula, we note that ${\rho}$ is a convex 
combination of mutually orthogonal one dimensional projectors, $P_{k}$, of its eigenvectors,  i.e.
$${\rho}={\sum}_{k}w_{j}P_{k}.\eqno(3.2)$$
Thus $w={\lbrace}w_{k}{\rbrace}$ is a probability measure on the pure states ${\lbrace}P_{k}{\rbrace}$ 
and Eq. (3.1) is equivalent to the formula
$$S({\rho})=-{\sum}_{k}w_{k}{\rm log}_{2}(w_{k}),\eqno(3.3)$$
which is just Shannon\rq s formula [SW] for the entropy of the probability measure $w$. Indeed, in the 
case where ${\Sigma}$ is a classical system and ${\rho}$ is a probability measure on a discrete space $K$, 
the entropy $S({\rho})$ is given by Eq. (3.3), with $w_{k}$ the probability attached to the pure state 
represented by the point $k$ of $K$.
\vskip 0.2cm
The formula (3.3), and hence also (3.1), has the natural interpretation [SW, Kh, Sz] that $-S({\rho})$ 
represents the information carried by the state ${\rho}$; or, equivalently, that the value of $S({\rho})$ is a 
measure of the degree of disorder of that state. To be precise, it is a strictly {\it probabilistic} measure of 
that disorder since , by Eq. (3.3), $S({\rho})$ depends exclusively on the probability measure $w$ and not 
at all on the structures of the pure states $P_{j}$.   
\vskip 0.2cm
Turning now to the case where ${\Sigma}$ is an infinite system, as formulated in Section 2, we denote by 
${\rho}_{\Lambda}$ the restriction of a state ${\rho}$ to the local algebra ${\cal A}_{\Lambda}$. The 
entropy density induced by ${\rho}$ in the region ${\Lambda}$ is therefore 
$S({\rho}_{\Lambda})/{\vert}{\Lambda}{\vert}$, where the numerator is the Von Neumann entropy of 
${\rho}_{\Lambda}$ and the denominator is the volume of ${\Lambda}$. It then follows from the strong 
subadditivity of entropy [LR] that, for any translationally invariant state ${\rho}$, this local entropy density  
converges to a limit $s({\rho})$ as ${\Lambda}$ increases to $X$ over a set of suitably regular regions, i.e. 
$${\rm lim}_{{\Lambda}{\uparrow}X}{S({\rho}_{\Lambda})\over {\vert}{\Lambda}{\vert}}=s({\rho}) \ 
{\forall} \ {\rho}{\in}{\cal S}_{X}.\eqno(3.4)$$
Evidently, it follows from the discussion following Eq. (3.3) that $s({\rho})$ is a strictly probabilistic 
measure of the disorder of the state ${\rho}$.
\vskip 0.3cm
{\bf Algorithmic Complexity and Disorder.} A complementary, intrinsic characterisation of disorder, as 
applied to pure states, has been provided by Kolmogorov [Ko] in the classical regime, and quantum 
versions of this have subsequently been proposed by Berthiaume et al [BVL], Gacs [Ga] and Vitanyi [Vi]. 
This is based on the concept of {\it algorithmic complexity}, which was introduced by Kolmogorov [Ko] in 
the context of classical communication theory, in the following form. For any string, $k_{N}$, of $N$ 
symbols, drawn from the binary set ${\lbrace}0,1{\rbrace}$, the algorithmic complexity $C(k_{N})$  is 
defined to be the length of the shortest programme required for the precise specification of that string by a 
universal Turing machine. Such a string of $0$\rq s and $1$\rq s corresponds to a pure state of a one-
dimensional classical lattice gas, ${\Sigma}_{N}$, whose phase space is 
$K_{N}:={\lbrace}0,1{\rbrace}^{[1,N]}$. 
\vskip 0.2cm
In order to pursue the properties of this algorithmic complexity, we treat ${\Sigma}_{N}$ as a subsystem 
of the infinitely extended classical lattice gas, ${\Sigma}$, whose phase space is 
$K:={\lbrace}0,1{\rbrace}^{\bf Z}$. Thus the elements of $K$ are the maps $k:x{\rightarrow}k_{x}$ of 
${\bf Z}$ into ${\lbrace}0,1{\rbrace}$ and the spatially ergodic states of ${\Sigma}$ are defined as in 
Section 2. The following theorem, due to Brudno [Br] provides a remarkable relationship between 
algorithmic complexity and entropy in terms of these definitions.
\vskip 0.3cm
{\bf Theorem 3.1} [Br]. {\it Let ${\rho}$ be a spatially ergodic measure on $K$ and, for $k{\in}K$, let 
$k_{N}$ be the restriction of $k$ to $[1,N]$. Then, for ${\rho}$-almost all $k{\in}K$, 
$${\rm lim}_{N\to\infty}N^{-1}C_{N}(k)=s({\rho}).\eqno(3.5)$$ }
\vskip 0.3cm
{\bf Comment.} This thereom signifies that, for ${\rho}$-almost all $k$ in $K$, the algorithmic 
complexity density $C(k_{N})/N$ induced by $k$ on the segment $[1,N]$ of ${\bf Z}$ converges to the 
entropy density $s({\rho})$ as $N{\rightarrow}{\infty}$. Hence, as algorithmic complexity is an intrinsic 
measure of disorder, the theorem vindicates the standard representation of the disorder of a macroscopic 
system by its entropy, at least in the case of one-dimensional lattice gases. From the physical standpoint, 
this does not conflict with the fact that the entropy of a pure state is zero, for the following reason. The 
specification of a pure state of a system with $N$ degrees of freedom would require the evaluation of $N$ 
variables. In the case of a macroscopic system, for which $N$ is extremely large (e.g. even in the case of a 
one-dimensional one, it is typically of the order of $10^{8}$), such a specification is out of the question. 
Indeed, for such a system, the only accessible information about its state is limited to the determination of a 
\lq few\rq macroscopic variables. The state inferred therefrom is then a highly mixed one. 
\vskip 0.3cm
{\bf Quantum Systems.} The above picture of entropy, complexity and disorder has been extended to 
quantum systems in the following way. Firstly, Berthiaume et al [BVL] have formulated  the algorithmic 
complexity of a pure state, $p$, of a string of $N$ qubits  as a natural quantum analogue of 
Kolmogorov\rq s classical picture (namely the length of the shortest programme required for the 
determination of $p$ by a universal quantum Turing machine) and Benatti et al [BKMSS] have established 
a quantum version of Brudno\rq s theorem for this complexity. These works were formulated within the 
framework of quantum informatics, but they may easily be translated into statistical mechanical terms by 
noting that a string of qubits corresponds to a system of Pauli spins on a one-dimensional lattice. The 
infinitely extended version, ${\Sigma}$, of such a system is thus a particular case of the quantum lattice 
model of Sec. 2 for which $d=1$ and the single particle Hilbert space ${\cal H}_{0}$ is two-dimensional. 
For any natural number $N$, we denote by ${\Sigma}_{N}$ the subsystem of ${\Sigma}$ comprising the 
spins at the sites $1,2, \ ,N$ and by 
${\cal A}_{N}$ its algebra of observables. For any state ${\rho}$ of ${\Sigma}$, we denote by 
${\rho}_{N}$ the state of ${\Sigma}_{N}$ given by its restriction to ${\cal A}_{N}$ and, following 
[BKMSS], we term a sequence of projectors ${\lbrace}P_{N}{\in}{\cal A}_{N}{\rbrace}$ {\it ${\rho}$-
typical} if ${\rm lim}_{N\to\infty}{\rho}(P_{N})=1$.Further, if $p_{N} \ ({\in}{\cal A}_{N})$ is a one-
dimensional projector we denote by $C(p_{N})$ the algorithmic complexity of this state of 
${\Sigma}_{N}$ that it represents, as formulated by [BVL]. The quantum version of Brudno\rq s theorem 
then takes the following form [BKMSS].
\vskip 0.3cm
{\bf Theorem 3.2.} {\it Let ${\rho}$ be a spatially ergodic state of the infinite chain, ${\Sigma}$, of Pauli 
spins. Then there exists a sequence of ${\rho}$-typical projectors 
${\lbrace}P_{N}{\in}{\cal A}_{N}{\rbrace}$ such that, for any ${\epsilon}>0$, every one-dimensional 
projector $p_{N}<P_{N}$ satisfies the following inequality for $N$ sufficiently large.
$$N^{-1}C(p_{N}){\in}\bigl(s({\rho})-{\epsilon},s({\rho})+{\epsilon}\bigr).\eqno(3.6)$$} 
\vskip 0.3cm
{\bf Comment.} This theorem signifies that the comment following Theorem 3.1 carries through to its 
natural quantum analogue. Further, as pointed out in [BKMSS], the above Theorem 3.2 is extendible to any 
chain of atoms, the observables of each of which are irreducibly represented in a finite dimensional Hilbert 
space ${\cal H}_{0}$. Moreover, we can extend that theorem to any $d$-dimensional lattice system, 
${\Sigma}$, as formulated in Section 2, in the following way. We replace the string $[1,N]$ of the one-
dimensional lattice by the block ${\Lambda}_{N}:=[1,N]^{d}$ of ${\bf Z}^{d}$ in the definitions of 
${\Sigma}_{N}, \ {\cal H}_{N}$ and ${\cal A}_{N}$, so that ${\Sigma}_{N}$ is now the system of 
atoms occupying the block ${\Lambda}_{N}$. The algorithmic complexity, $C(p_{N})$ of a pure state 
$p_{N}$ of this system is then the length of the shortest programme required to specify this state by a 
universal quantum Turing machine and its complexity density is $C(p_{N})/N^{d}$. Further, if ${\rho}$ 
is a state of the infinite system ${\Sigma}$, a sequence of projectors ${\lbrace}P_{N}{\in}{\cal 
A}_{N}{\rbrace}$ is again termed  ${\rho}$-typical if ${\rm lim}_{N\to\infty}{\rho}(P_{N})=1$.With 
these definitions, the treatment of [BKMSS] can be carried through to yield the following $d$-dimensional 
generalisation of Theorem 3.2. 
\vskip 0.3 cm
{\bf Theorem 3.3.} {\it Let ${\rho}$ be a spatially ergodic state of the infinite system, ${\Sigma}$, of 
atoms on the lattice ${\bf Z}^{d}$. Then there exists a ${\rho}$-typical sequence of projectors 
${\lbrace}P_{N}{\in}{\cal A}_{N}{\rbrace}$ such that, for any ${\epsilon}>0$, the algorithmic  
complexity, $C(p_{N})$, of any one-dimensional  projector $p_{N}<P_{N}$ satisfies the following 
inequality for $N$ sufficiently large. 
$$N^{-d}C(p_{N}){\in}\bigl(s({\rho})-{\epsilon},s({\rho})+{\epsilon}\bigr).\eqno(3.7)$$} 
\vskip 0.3cm
{\bf Comment.} This theorem signifies that the comments following Theorems 3.1 and 3.2 extend to 
quantum systems on lattices of arbitrary finite dimensionality and thus vindicates the standard picture 
wherein the disorder of a state of a quantum lattice system is given by its Von Neumann entropy.
\vskip 0.5cm
\centerline {\bf 4. Symmetry and Order.}
\vskip 0.3cm
{\bf Symmetry Groups and $G$-fields.} A {\it symmetry group} of the model ${\Sigma}=({\cal A},{\cal 
S}, {\gamma},{\alpha})$ is a group, $G$, that has a faithful representation, ${\theta}$, in ${\rm Aut}({\cal 
A})$. It is a {\it dynamical symmetry group} if ${\theta}(G)$ commutes with the time-translations 
${\alpha}_{t}$. For any symmetry group $G$, an $n$-component quantum field ${\xi}= ({\xi}_{1},. \ 
.,{\xi}_{n})$, affiliated to ${\cal A}$, is termed a {\it $G$-field} if  the action of ${\theta}(G)$ on ${\xi}$ 
takes one of the following forms.
\vskip 0.2cm\noindent
(a) In the case where this symmetry group is spatial,
$$[{\theta}(g){\xi}](x)={\xi}(T_{g}x) \ {\forall} \ g{\in}G,\eqno(4.1a)$$
where $T_{g}$ is a transformation of $X$.
\vskip 0.2cm\noindent
(b) In the case where $G$ is an internal symmetry group,
$$[{\theta}(g){\xi}]_{j}(x)={\sum}_{k=1}^{n}V_{g;jk}{\xi}_{k}(x) \ {\forall} \ g{\in}G,\eqno(4.1b)$$ 
where $V_{g}=[V_{g,jk}]$ is a unitary transformation of ${\bf R}^{n}$ or ${\bf C}^{n}$, according to 
whether the field ${\xi}$ is real or complex..
\vskip 0.2cm\noindent
Thus, in either case, the action of ${\theta}(g)$ on ${\xi}$ is that of a linear transformation ${\phi}_{g}$ 
of this field alone, without involvement of other observables of the system, i.e. 
$$[{\theta}(g){\xi}](x)=[{\phi}_{g}{\xi}](x):={\xi}_{g}(x).\eqno(4.1)$$
\vskip 0.3cm
{\bf Symmetry Breakdown.} An stationary state ${\rho}$ of ${\Sigma}$ that is not invariant with respect 
to a dynamical symmetry group $G$ is said to spontaneously break that symmetry. In this case, the states 
${\lbrace}{\rho}_{g}:={\rho}{\circ}{\theta}(g){\vert}g{\in}G{\rbrace}$ are also $G$-symmetry breaking 
stationary states of the system. We term this set of states the $G$-orbit of ${\rho}$ and denote it by ${\cal 
O}_{G}({\rho})$.
\vskip 0.2cm
 In the case where ${\Sigma}$ is conservative and ${\rho}{\in}{\cal E}({\cal S}_{\beta})$, the set of 
extremal KMS states at inverse temperature ${\beta}$, it follows from the KMS condition (2.1) and the 
definition of ${\cal O}_{G}({\rho})$ that ${\cal E}({\cal S}_{\beta}){\subset}{\cal O}_{G}({\rho})$. 
Further, defining 
$${\overline {\xi}}(x):= {\langle}{\rho};{\xi}(x){\rangle} \ {\rm and} \ 
{\overline {\xi}}_{g}(x):= {\langle}{\rho}_{g};{\xi}(x){\rangle},\eqno(4.2)$$
it follows from Eq. (4.1) that
$${\overline {\xi}}_{g}(x) =[{\phi}_{g}{\overline {\xi}}](x),\eqno(4.3)$$
where the action of ${\phi}_{g}$ on the classical field ${\overline {\xi}}$ is the same as that on the 
quantum field ${\xi}$. We assume that this action is non-trivial\footnote*{In fact, this condition is 
effectively fulfilled if the subgroup, $H$, of $G$ under which the field ${\overline {\xi}}$ is invariant is a 
normal one. For in that case we may replace the group $G$ of the present treatment by the factor group 
$G/H$, which then satisfies our demands.}  unless $g$ is the identity element of $G$ and consequently that 
${\overline {\xi}}_{g}{\neq}{\overline {\xi}}_{g^{\prime}}$ if $g{\neq}g^{\prime}$. Hence, by Eqs. 
(4.1)-(4.3), the $G$-field ${\xi}$ serves to separate the states of the orbit ${\cal O}_{G}({\rho})$. We note 
here that, since these states are primary, it follows from Eqs. (2.4) and (4.1)-(4.3) that 
$${\rm lim}_{a\to\infty}[{\langle}{\rho};{\xi}_{g}^{\star}(x).{\xi}_{g}(x+a){\rangle}-
{\overline {\xi}}_{g}^{\star}(x).{\overline {\xi}}_{g}(x+a)]=0,\eqno(4.4)$$
where the dot denotes the scalar product in ${\bf R}^{n}$ or ${\bf C}^{n}$.
\vskip 0.3cm
{\bf Symmetry Breakdown and Order.} At a qualitative level, we conceive a state of a complex system to 
be ordered if its components cooperate in such a way as to produce a macroscopic field or signal. Thus, a 
prototype example of ordering in this sense is that of the alignment of the spins (mini-magnets!) of a 
ferromagnetic material so as to produce a resultant polarisation. Hence, assuming that the dynamics of the 
material is rotationally invariant, this magnetic ordering amounts to a breakdown of its rotational 
symmetry, as manifested by the direction of the polarisation field.
\vskip 0.2cm
The generalisation of this concept of order through symmetry breakdown is straightforward. Thus, an 
equilibrium state ${\rho}$ of ${\Sigma}$ is said to be ordered if it is not invariant with respect to a 
dynamical symmetry group $G$, and its ordering is then represented by the $g$-dependence of the 
classical field ${\overline {\xi}}_{g}$. Among the numerous proven examples of such order are the 
following.
\vskip 0.2cm\noindent
{\it Example 1.} The two-dimensional Ising model, whose pure equilibrium phases at any temperature 
below its critical point are polarised [MM]. For this model, $G$ is the binary group $(e,r)$, where $e$ is 
the identity element and  ${\theta}(r)$ is the spin reversal automorphism. The $G$-field ${\xi}(x)$ is the 
Ising spin $s_{x}$ at the site $x$. 
\vskip 0.2cm\noindent
{\it Example 2.} The antiferromagnetic phase of the Heisenberg model [DLS]. For this, $G$ is the three 
dimensional rotation group and ${\theta}(G)$ represents its action on the Pauli spins constituting the 
model. The $G$  field ${\xi}(x)$ is the Pauli spin vector ${\sigma}(x)$ at the site $x$ and so ${\overline 
{\xi}}_{g}(x)$ is a spatially periodic vector filed whose direction is determined by $g$.
\vskip 0.2cm\noindent
{\it Example 3.} A crystalline phase of the generic model of a continuous system ${\Sigma}$ that occupies 
the space $X={\bf R}^{d}$. For this system, $G$ is the factor group $X/Y$, where $X$ is the additive 
space translation group and $Y$ is the normal subgroup of $X$ corresponding to the crystal structure 
[EKV]. The $G$-field ${\xi}(x)$ may be chosen to be the particle density ${\psi}^{\star}(x){\psi}(x)$.
\vskip 0.2cm\noindent
{\it Example 4.} This is the order corresponding to a generalised version of Bose-Einstein (BE) 
condensation, which was first proposed by O. Penrose and Onsager [PO] as a characterisation of the 
superfluidityof HeII  and subsequently extended by Yang [Ya] to superconductors. In fact, this order 
corresponds to the breakdown of gauge symmetry\footnote*{See [Se1, Ch.9] for a detailed discussion of 
this symmetry breakdown and its connections with both superfluidity and the so-called \lq off-diagonal 
long range order\rq\ of  O. Penrose and Onsager [OP] and Yang [Ya].}, the relevant dynamical symmetry 
group, $G$, being that of the gauge transformations ${\psi}(x){\rightarrow}{\psi}(x)({\rm exp}(ic)$, 
where $c$ runs through the reals. In the case of bosons, the $G$-field ${\xi}(x)$ is just ${\psi}(x)$: in the 
case of fermions it is the pair field ${\psi}_{\uparrow}(x){\psi}_{\downarrow}(x)$, the two factors being 
the components of ${\psi}(x)$ with spin parallel and antiparallel to some fixed axis.
\vskip 0.2cm\noindent
{\it Example 5.} This is the pumped phonon model, which was proposed by H. Froehlich [Fr] in a 
biological context and put onto a rigorous footing by Duffield [Du]. It is an open dissative system, 
consisting of $N$ phonon modes that are coupled to energy pumps and sinks and that exchange quanta with 
one another in conformity with the principle of detailed balance at the prevailing environmental 
temperature. Remarkably, the model is driven by these forces into a state wherein a macroscopic number, 
of order $N$, of its quanta condense into the mode of lowest frequency when the pumping strength exceeds 
a critical value. Thus, in the limit $N{\rightarrow}{\infty}$, it exhibits a BE condensation into a 
nonequilibrium steady state. This amounts to a gauge symmetry breakdown, far from thermal equilibrium. 
The source of this phenomenon stems is the competition between the pumping and the discharge of quanta 
into the sinks, which fixes the total number of quanta and thereby renders the state of the model similar to 
an equilibrium state of an ideal Bose gas with a fixed number of particles. By contrast, a phonon system at 
equilibrium with a thermostat at fixed temperature has a Planck distribution of its quanta and thus does not 
experience a BE condensation.
\vskip 0.5cm
\centerline {\bf 5. Coherence.}
\vskip 0.3cm
Coherence is an extreme version of order. Specifically, a state ${\rho}$ of the system ${\Sigma}$ is said to 
be coherent with repect to a $G$-field, ${\xi}(x)$, if it satisfies Glauber\rq s [Gl] condition that 
$${\langle}{\rho};{\xi}^{\#}(x_{1}). \ .{\xi}^{\#}(x_{n}){\rangle}= {\Pi}_{j=1}^{n} 
{\langle}{\rho}:{\xi}^{\#}(x_{j}){\rangle},\eqno(5.1)$$ 
where each ${\xi}^{\#}$ is either ${\xi}$ or ${\xi}^{\star}$ and ${\langle}{\rho};{\xi}(x){\rangle}$ is a 
non-trivial function of $x$, which is simple in that it involves just a \lq few\rq\  parameters: the simplicity 
condition serves to exclude cases where the classical field represented by this function varies chaotically 
with $x$. 
\vskip 0.2cm
The natural dynamical version of this coherence condition is that obtained by the replacement, in Eq. (5.1), 
of ${\xi}(x)$ by its evolute ${\xi}_{t}(x) \ (={\alpha}_{t}{\xi}(x)$). Thus, the resultant dynamical 
coherence codition is that
$${\langle}{\rho};{\xi}_{t_{1}}^{\#}(x_{1}). \ .{\xi}_{t_{n}}^{\#}(x_{n}){\rangle}= {\Pi}_{j=1}^{n} 
{\langle}{\rho}:{\xi}_{t_{j}}^{\#}(x_{j}){\rangle},\eqno(5.2)$$ 
where ${\langle}{\rho};{\xi}_{t}(x){\rangle}$ is a non-trivial, simple function of both $x$ and $t$.
\vskip 0.2cm
It follows from these specifications that coherence corresponds to the behaviour of the quantum field 
${\xi}(x)$ or ${\xi}_{t}(x)$, in the state ${\rho}$, as a classical, dispersion-free field 
${\overline {\xi}}(x):={\langle}{\rho};{\xi}(x){\rangle}$ or 
${\overline {\xi}}_{t}(x):={\langle}{\rho};{\xi}_{t}(x){\rangle}$
\vskip 0.2cm
The following examples of coherence have been established in different variants of the Dicke model [Di] of 
matter interacting with a single radiative mode.
\vskip 0.3cm
{\it Example 1.} Hepp and Lieb [HL2] have proved that the Dicke model has a low temperature 
equilibrium phase characterised by super-radiance, i.e. by coherence of the pure equilibrium phase with 
respect to the radiation field, with breakdown of the gauge symmetry of that field. They also treated [HL1] 
the {\it open} version of the Dicke model in which each atom of the matter was coupled to an energy pump 
and a sink and the radiation mode was coupled to a sink. This model undegoes a  phase transition, far from 
thermal equilibrium, at a critical value, $p_{c}$, of the pumping strength, $p$. Specifically, for 
$p<p_{c}$, its  stable state is stationary and its radiation incoherent; while for $p>p_{c}$, the stable state 
is simply periodic in time and its radiative mode coherent, again with breakdown of gauge symmetry. Thus 
the transition is from normal light to laser light. 
\vskip 0.3cm
{\it Example 2.} Alli and Sewell [AS] extended the open version of the Dicke model to one with many 
modes, each with its own sink, and obtained the following picture of its phase structure in terms two critical 
values, $p_{1}$ and $p_{2} \ (>p_{1})$ of the pumping strength. For $p<p_{1}$, there is a unique stable 
stationary state of the model, and the radiation is normal, i.e. incoherent. For $p_{1}<p<p_{2}$, the stable 
state varies periodically with time and  the radiation is coherent and monochromatic, again with breakdown 
of gauge symmetry. For $p>p_{2}$, the radiation is chaotic according to either the mechanism of Ruelle-
Takens [RT] (strange attractor) or that of Landau [LL] (multimode turbulence), depending on the 
parameters of the model.
\vskip 0.3cm
{\it Example 3.} Pule, Verbeure and Zagrebnov [PVZ] constructed a model of a system of interacting two 
level bosonic atoms that is coupled to a radiative mode. This coupling was shown to leads to a rich 
equilibrium phase structure. In particular, for sufficiently large values of the chemical potential, the 
radiation is coherent and the matter exhibits a two-fold BE condensation, one for each of its atomic levels. 
This supports the experimental observation that the action of a laser field on a bosonic condensate of atoms 
with internal structure leads to enhancements of both the laser field and the BE condensation [KI].
\vskip 0.5cm
\centerline {\bf 6. Concluding Remarks} 
\vskip 0.3cm
We may summarise the contents of this article as follows. 
\vskip 0.2cm
Disorder corresponds to randomness, of which entropy and algorithmic complexity provide probabilistic 
and intrinsic measures, respectively. Remarkably, Brudno\rq s theorem and its quantum analogues have 
established the essential equivalence of these measures in the thermodynamic limit (cf. Theorems 3.1-3.3). 
\vskip 0.2cm
Order, on the other hand, constitutes organisation manifested by a macroscopic field or signal, and this can 
prevail amidst high disorder. Thus, in the present physical context, the concepts of order and disorder are 
certainly not antitheses of one another. The particular types of order and coherence described here stem 
from symmetry breakdown and BE condensation, and certainly do not cover all the kinds of organisation 
that arise in complex systems. For example, it is still true that, as pointed out by Schroedinger many years 
ago [Sc], the presently available pictures of order do not cover biological organisation..Thus, a major 
challenge of statistical physics is to characterise other kinds of ordering that exist in nature. 
\vskip 0.5cm
{\bf Acknowledgement.} It is a pleasure to thank Fabio Benatti for an enlightening correspondence on the 
subject of algorithmic complexity.
\vskip 0.5cm
\centerline {\bf References}
\vskip 0.3cm\noindent
[AS] G. Alli and G. L. Sewell: J. Math. Phys. {\bf 36}, 5598-5626, 1995.
\vskip 0.2cm\noindent
[Br] A. A. Brudno: Trans. Moscow Math. Soc. {\bf 2}, 127-151, 1983.
\vskip 0.2cm\noindent
[BCS] J. Bardeen, L. N. Cooper and J. R. Schrieffer: Phys. Rev. {\bf 108}, 1175-1204, 1957.
\vskip 0.2cm\noindent
[BVL]  A. Berthiaume, W. Van Dam and S. Laplante: J. Comput. System Sci. {\bf 63}, 201-221, 2001.
\vskip 0.2cm\noindent
[BKMSS] F. Benatti, T. Kruger, M. Mueller, R. Siegmund-Schultze and A. Szkola: Commun. Math. Phys. 
{\bf 265}, 437-461, 2006.
\vskip 0.2cm\noindent
 [Di]  R. H. Dicke: Phys. Rev. {\bf 93}, 99-110, 1954.
\vskip 0.2cm\noindent
[Du] N. G. Duffield: Phys. Lett. {\bf 110A}, 332-334, 1985; J. Phys. A {\bf 21}, 625-641, 1988.
\vskip 0.2cm\noindent
[DDR] G.-F. Dell\rq\ Antonio, S. Doplicher and D. Ruelle: Commun. Math. Phys. {\bf 2}, 223-230, 1966.
\vskip 0.2cm\noindent
[DLS]  F. J. Dyson, E. H. Lieb and B. Simon: J. Stat. Phys. {\bf 18}, 335-383, 1978.
\vskip 0.2cm\noindent
[Em]  G. G. Emch: {\it Algebraic Methods in Statistical mechanics and Quantum Field Theory}, Wiley, 
New York, 1972.
\vskip 0.2cm\noindent
[EKV] G. G. Emch, H. J. F. Knops and E. Verboven: J. Math. Phys. {\bf 11}, 1655-1667, 1970. 
\vskip 0.2cm\noindent
[Fe]  R. P. Feynman: Prog. Low Temp. Physics {\bf 1}, 17-53, 1955.
\vskip 0.2cm\noindent
[Fr] H. Froehlich: Int. J. Quantum Chem. {\bf 2}, 641-649, 1968.
\vskip 0.2cm\noindent
[Ga] P.gacs: J. Phys. A {\bf 34}, 6859-6880, 2001.
\vskip 0.2cm\noindent
[Gl] R. J. Glauber: Phys. Rev. {\bf 130}, 2529-2539, 1963.
\vskip 0.2cm\noindent
[HL1] K. Hepp and E. H. Lieb: Helv. Phys. Acta {\bf 46}, 573-603, 1973.
\vskip 0.2cm\noindent
[HL2] K. Hepp and E. H. Lieb: Ann. Phys {\bf 76}, 360-404, 1973
\vskip 0.2cm\noindent                                           
[HHW] R. Haag, N. M. Hugenholtz and M. Winnink: Commun. Math. Phys. {\bf  5}, 215-236, 1967.
\vskip 0.2cm\noindent
[Jo]  B. D. Josephson: Rev. Mod. Phys. {\bf 36}, 216-220, 1964.
\vskip 0.2cm\noindent
[Kh] A. I. Khinchin: {\it Mathematical Foundations of Information Theory}, Dover, New York, 1971.
\vskip 0.2cm\noindent
[Ko] A. N. Kolmogorov: IEEE Trans. Inform. Theory {\bf 14}, 662-664, 1968.
\vskip 0.2cm\noindent
[KI] W. Ketterle and S. Inouye: C. R. Acad. Sci. Paris Serie IV {\bf 2}, 339-380, 2001.
\vskip 0.2cm\noindent
[Li] G. Lindblad: Commun. Math. Phys. {\bf 48}, 119-130, 1976.
\vskip 0.2cm\noindent
[LL] L. D. Landau and E. M. Lifshitz: {\it Fluid Mechanics}, Pergamon, Oxford, 1984.
\vskip 0.2cm\noindent
[LR] E. H. Lieb and M. B. Ruskai: J. Math. Phys. {\bf 14}, 1938-1941, 1973.
\vskip 0.2cm\noindent
[LSSY] E. H. Lieb, R. Seiringer, J. Solovej and J. Yangvason: {\it The Mathematics of the Bose Gas and its 
Condensation}, Birkhauser, berlin, 2005.
\vskip 0.2cm\noindent
[MM]  A. Messager and S. Miracle-Sole: Commun. Math. Phys. {\bf 40}, 187-198, 1975.
\vskip 0.2cm\noindent
[PO] O.Penrose and L. Onsager: Phys. Rev. {\bf 104}, 576-584, 1956.
\vskip 0.2cm\noindent
[PVZ] J. Pule, A. Verbeure and V. Zagrebnov: J. Stat. Phys. {\bf 119}, 309-329, 2005.
\vskip 0.2cm\noindent
[Ro] D. W. Robinson: Commun. Math. Phys. {\bf 7}, 337-348, 1968.
\vskip 0.2cm\noindent
 [Ru] D. Ruelle: Pp. 169-194  of {\it Cargese Lecures}, Vol. 4, Ed. D. Kastler, Gordon and Breach, New 
York, 1969.
\vskip 0.2cm\noindent
[RT] D. Ruelle and F. Takens: Commun. Math. Phys. {\bf 20}, 187-192, 1971.
\vskip 0.2cm\noindent 
[Sc] E. Schroedinger: {\it What is Life?}, cambridge Univ. Press, 1967. 
\vskip 0.2cm\noindent
[Se1] G. L. Sewell: {\it Quantum Mechanics and its Emergent Macrophysics}, Princeton Univ. Press, 
Princeton, 2002.
\vskip 0.2cm\noindent
[Se2] G. L. Sewell: Lett. Math. Phys. {\bf 6}, 209-213, 1982.
\vskip 0.2cm\noindent
[St] R. F. Streater: Commun. Math. Phys. {\bf 6}, 233-247, 1967.
\vskip 0.2cm\noindent
[Sz] L.Szilard: Z. Phys. {\bf 32}, 777, 1925.
\vskip 0.2cm\noindent
[SW] C. E. Shannon and W. Weaver: {\it The Mathematical Theory of Communication}, Illinois Univ. 
Press, IL, 1949.
\vskip 0.2cm\noindent
[Th] W. Thirring: {\it Quantum Mechanics of Large Systems}, Springer, New York, 1983.
\vskip 0.2cm\noindent
[TW] M. Takesaki and M. Winnink: Commun. Math. Phys. {\bf 30}, 129-152, 1973.
\vskip 0.2cm\noindent 
[Vi] P. Vitanyi: IEEE Trans. Inform. Theory {\bf 47/6}, 2464-2479, 2001.
\vskip 0.2cm\noindent
[VN] J. Von Neumann: {\it Mathematical Foundations of Quantum Mechanics}, Princeton Univ. Press, 
Princeton, 1955.
\vskip 0.2cm\noindent
[We] A. Wehrl: Rev. Mod. Phys. {\bf 50}, 221-260, 1978.
\vskip 0.2cm\noindent
[Ya] C. N. Yang: Rev. Mod. Phys. {\bf 34}, 694-704, 1962
\vskip 0.2cm\noindent

\end